\documentclass[12 pt]{article}

\usepackage[utf8]{inputenc}
\usepackage{amsthm}
\usepackage{amssymb}
\usepackage{amsmath}
\usepackage{amsfonts}
\usepackage{geometry}
\usepackage{commath}
\usepackage{graphicx}
\usepackage{subcaption}
\usepackage{bm}
\usepackage{float}
\usepackage{caption}
\usepackage{setspace}
\usepackage{multicol}
\usepackage{color}
\usepackage{subfiles}
\usepackage{textcomp}
\usepackage{hyperref}
\usepackage{commath}
\usepackage{mathrsfs}
\usepackage{listings}
\usepackage{gensymb}
\usepackage{empheq}
\usepackage{minted}
\usepackage{xcolor}
\usepackage[title]{appendix}
\usepackage{authblk}
\usepackage{natbib}
\usepackage{lineno}
\usepackage{indentfirst}

\newtheorem*{definition*}{Definition}

\newtheorem*{theorem*}{Theorem}

\newtheorem*{corollary*}{Corollary}

\newtheorem*{solution*}{Solution}

\newtheorem*{example*}{Example}

\title{An Integrated Time-Varying Ornstein-Uhlenbeck Process for Jointly Modeling Individual and Population-Level Movement of Golden Eagles}
\author[a]{Michael L. Shull\footnote{Corresponding author - email: mlshull@psu.edu}}
\author[a]{Ephraim M. Hanks}
\author[b]{James C. Russell}
\author[c]{Robert K. Murphy}
\author[d]{Frances E. Buderman}
\affil[a]{Department of Statistics, The Pennsylvania State University, University Park, PA, USA}
\affil[b]{Department of Mathematics, Computer Science, and Statistics, Muhlenberg College, Allentown, PA, USA}
\affil[c]{Eagle Environmental Inc., Santa Fe, New Mexico, USA}
\affil[d]{Department of Ecosystem Science and Management, The Pennsylvania State University, University Park, PA, USA}
\date{}

\begin{document}

\maketitle

\begin{abstract}
With technological advancements, the quantity and quality of animal movement data have increased greatly. Currently, no movement model can be used to describe full-year data from migratory species by leveraging both individual movement and species distribution data. Herein we propose a full-year stochastic differential equation model for jointly modeling both individual movement and species distribution data.  We show that this joint model, under certain assumptions, results in efficient computation of the spatio-temporal dynamics of the entire population, and thus provides straightforward inference on the species distribution data. We illustrate this model by analyzing 215 bird-years of golden eagle movement in western North America jointly with relative abundance data from eBird. We use the results to estimate wind project risk for these eagles and predict where they came from earlier in the year based on a single telemetry observation from later in the year. Our joint model enables additional inference and greater predictive power than afforded by sole use of eBird relative abundance.

\end{abstract}

\section{Introduction}

Managing migratory species is challenging as it requires an understanding of spatio-temporal migration patterns, which can span nations and even continents. An understanding of these migration patterns can, for example, enable a better understanding of effects of climate and land-use change on migratory species, such as golden eagles (\textit{Aquila chrysaetos}) or help identify which government agencies should work in concert to manage migratory game species such as mallards (\textit{Anas platyrhynchos}). One land-use change that poses a direct risk to golden eagles is the development of wind energy projects \citep{Beston_2016}; golden eagle mortality due to collisions with wind turbines, specifically with spinning turbine blades \citep{hunt_2002, hunt2016addressing}, has increased steadily in the western U.S. and may have a noticeable population-level impact \citep{gedir_2025}. Knowing the spatio-temporal dynamics and distribution of eagles that are vulnerable to impact of current and future wind projects is of interest to government and private entities who perform environmental impact studies or are directly charged with stewardship of the eagle resource. These studies can inform appropriate placement of wind projects to mitigate the risk to golden eagles and other migratory species. For mallards, summer breeding grounds and winter ranges are often thousands of miles apart; therefore, government agencies that manage breeding grounds must work together with those that manage winter ranges to successfully maintain mallard populations \citep{Waterfowl}. To facilitate this cooperation we need to understand the spatio-temporal migration patterns so that we can estimate which breeding and winter grounds share common populations. Therefore, models that can explain the spatio-temporal dynamics of migratory species for the full-annual cycle are essential for principled management of migratory species like mallards and golden eagles.

One key source of data for migratory species is GPS derived telemetry data. These data provide time-indexed spatial information on how individuals in the population of interest move during both migratory and non-migratory seasons. However, telemetry data are inherently individual, but management goals are population-based. Using GPS data for population level inference should be done in a principled manner \citep{UD_from_CTMCM_wilson_2018,ScaleUp_StepSelection_potts_2023, winter_2024}, leveraging the heterogeneity among individuals \citep{IMM_original, Michelot_hanks_2025}. However, GPS data do not provide unbiased information regarding the spatio-temporal distribution of the entire population, as GPS tags are almost never distributed in a way that is representative of the entire population of interest \citep{gow2019effects,hertel2020guide,norevik2025spatial}; therefore, additional information is needed. Some studies attempt to address this issue by pooling many telemetry studies to garner a broader picture of the entire population \citep[e.g.][]{eagles_multiplestudies_2017}. However, this still relies on an assumption that individuals in the collective telemetry studies are representative of the entire population. One solution to this problem of generalizing from individual telemetry data is to include spatial survey data \citep[e.g.][]{FlyoverMethods_Neilson_2014, Flyoverwithtelemetry_Neilson_2016} and build a joint model for both types of data \citep{Blackwell_Matthiopoulos_telem_withsurvey_2024, telemwithsurvey_lauret_2025}. However, spatial surveys with no temporal component are not as informative in the case of migratory species where we want to generalize not based on a single point in time but throughout the whole year.

An additional source of movement data for migratory species comes from citizen science initiatives. For example, birdwatchers can report geo-referenced auditory or visual detections of individual birds to the eBird platform \citep{eBird_Sullivan2009}. Although some detected birds may be marked with bands or GPS trackers, most are unmarked. These records can then be used to produce estimates of the spatio-temporal distribution for a given species at a fine spatial resolution \citep{Fink_2013_adaSTEM}. This aggregated data product can approximate the spatio-temporal movement dynamics of a migratory species, similar to what we might expect with a time series of survey data; however, this population-level data source provides no insight into individual migration patterns. One recent approach of jointly modeling these two types of data (telemetry and eBird adaSTEM output) is the Integrated Movement Modeling (IMM) framework of \citet{IMM_original}. This method integrates a statistical model for individual telemetry data over a model for the heterogeneity of movement behavior within a population to obtain a probabilistic model for the spatio-temporal dynamics of the species' population. \citet{IMM_original} illustrate this framework in modeling golden eagle migration in western North America during 8 weeks of the spring migration.

In this paper, we build a joint full-annual-cycle model for these two data types to estimate full-annual-cycle migratory dynamics and assess the varying spatial risk from hazards to different subsets of the population during both migratory and non-migratory seasons. In this way we leverage individual migration patterns from GPS tracking data with information from citizen science data about the spatio-temporal distribution of the species. To this end, we develop a full-year time-varying Ornstein-Uhlenbeck (OU) model for animal movement, scale it up to derive a formal probabilistic model for the population-level movement, and show that this stochastic differential equation (SDE)  has an analytic solution that makes computing and inference very efficient, even at the continental-scale. This model we propose allows us to answer both population and individual-level questions about golden eagle movement behaviors. We use this model to examine how migration timing varies among individual eagles in the population and to estimate which wind projects in the contiguous western United States pose the greatest risk to golden eagles.

\section{Data} \label{sec:data}

We obtained movement data from golden eagles tracked by the U.S. Fish and Wildlife Service (USFWS) via GPS satellite telemetry. Eagles were tagged in the Central Great Plains, Colorado Plateau, Rocky Mountain (south of Montana), Texas Trans-Pecos, and Southern Great Plains regions; collectively, these regions encompass roughly half of the species' range in the coterminous western U.S. The majority (77.9\%) of the eagles were tagged with satellite transmitters when they were large (between 7 and 8 weeks old) nestlings which subsequently dispersed from their natal area by the end of their first year of life \citep{murhpy_2017_first}. The remaining eagles (22.1\%) were trapped and tagged when they were more than a year old, including some adults \citep[individuals more than four years old;][]{Murphy_2019}. The transmitters (solar Argos/GPS 45-g and 70-g platform terminal transmitter units; Microwave Telemetry, Inc., Columbia, MD) were attached in a backpack configuration via a “Y-harness” constructed of Teflon ribbon and collected GPS locations hourly each day during at least 0900-1600 H; transmitter location accuracy was $\pm$ 19 m \citep{murhpy_2017_first}. 

\begin{figure}[tbp]
    \centering
    \includegraphics[width = 5.61893in]{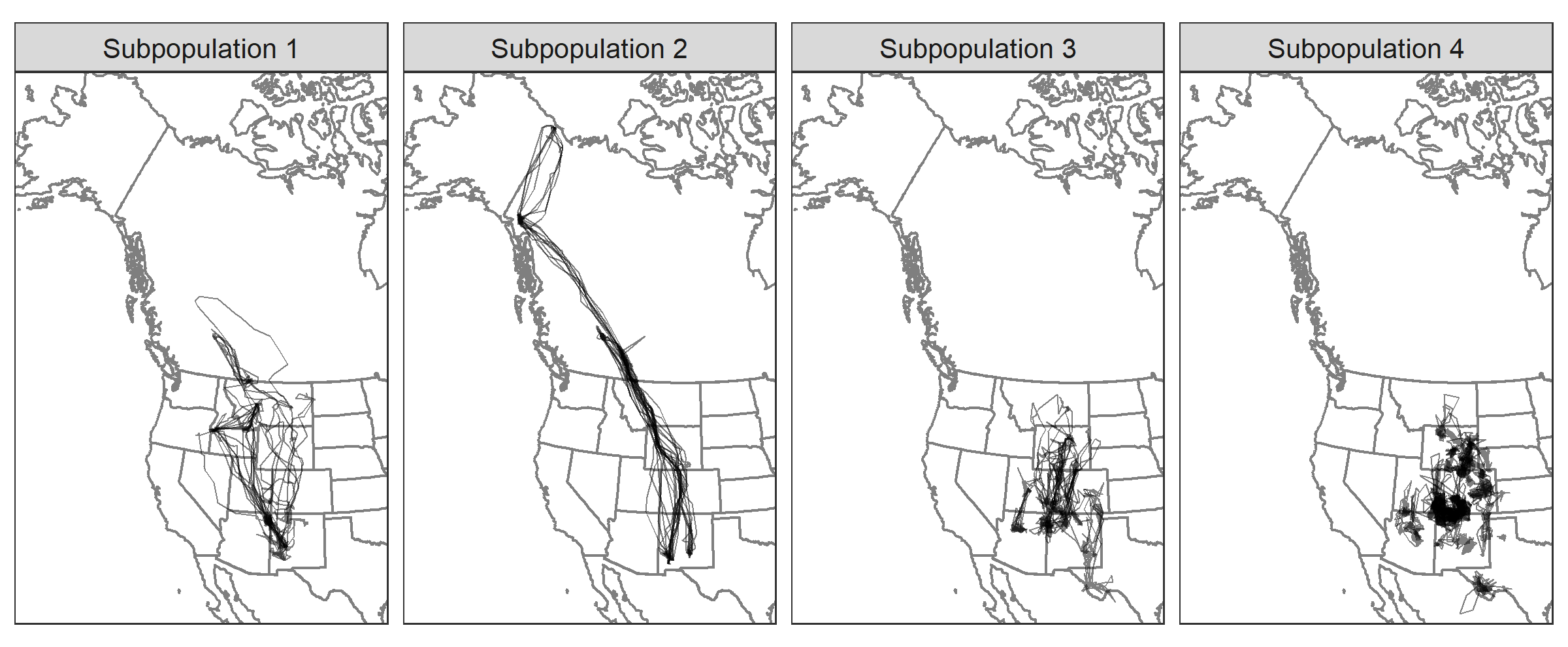}
    \caption{Movement tracks based on satellite telemetry data for golden eagles in western North America, separated by subpopulation. The eagles were placed in subpopulations based on a clustering of their normalized (displacement from January 1 location) telemetry data off via k-means algorithm. This highlights the different migratory behaviors that individual golden eagles exhibit. These four subpopulations could be described as moderate-distance migrants, long-distance migrants, short-distance migrants, and non-migratory (full-year resident) eagles respectively.}
    \label{fig:telemdata}
\end{figure}

In our study we use telemetry data from calendar years in which, for a given eagle, 315 days (including both January 1 and December 31) had at least one telemetry observation; this resulted in 215 calendar years of data from 93 golden eagles during 2011-2018; individual eagles in our sample were each represented by data from one to six complete years. Full information on which years were used for which golden eagles is available in Sections 1 and 6 of the Supplemental Material \citep{FullYearOU_Supp}. Our dataset for a given eagle included a single daily location for each 24-hour period, derived by averaging all GPS locations available for the 24-hour period. As we describe in Section \ref{sec:heterogeneity}, heterogeneity in movement behavior will be modeled via a mixture model. This is heuristically equivalent to the population being composed of several subpopulations. We cluster our 215 years of telemetry data into four groups using the k-means algorithm \citep{steinley_2006_kmeans}, with four groups chosen based on an examination of where the decrease in total within-cluster variation started to slow. The clustering was based on a vector of normalized telemetry data, where for each bird-year of data the individual's location on January 1 was subtracted from all the other telemetry readings for that individual (i.e. clustered based on displacement from the January 1 location). This was done so the subpopulations would be based on migratory behavior rather than simply the January 1 location. We assumed that eagles in these four subpopulations would differ in their geographic centroids, attraction to their respective centroids, and migration timing (these three characteristics comprise the parameters in our selected movement model). This procedure resulted in 11, 13, 24, and 167 bird-years in each subpopulation respectively. Figure \ref{fig:telemdata} shows the telemetry data divided by subpopulation. These four subpopulations could be described as moderate-distance migrants, long-distance migrants, short-distance migrants, and non-migratory (full-year resident) eagles respectively. 

For species distribution data, we use eBird adaSTEM output. The Cornell Lab of Ornithology has developed the Adaptive Spatio-Temporal Exploratory Model \citep[adaSTEM;][]{Fink_2013_adaSTEM, Fink_2014_crowdsourcing}, which processes huge numbers of individual citizen-science records and accounts for spatial heterogeneity in sampling effort, observer skill, and rarity of species to estimate the relative abundance of a species over space and time \citep{Fink_2013_adaSTEM}. The Cornell Lab of Ornithology defines relative abundance as "the count of individuals of a given species detected by an expert eBirder on a 1 hour, 2 kilometer traveling checklist at the optimal time of day" (\url{https://science.ebird.org/en/status-and-trends/faq#abundance}). Several studies have shown that relative abundances produced by the adaSTEM accurately represent avian spatial distributions \citep{Ruiz‐Gutierrez_2021, howell_2022, Stuber_2022, stillman_2023}. This eBird adaSTEM output information is available at a 2.8 x 2.8km resolution, at weekly intervals through a given year, throughout the golden eagles' range. 

We subset the 2018 relative abundance \citep{fink2021} down to western North America, aggregated to 150-km resolution, and normalized the gridded relative abundance to sum to unity for each time point. We used this weekly, normalized, eBird adaSTEM output as the species distribution data  in our analysis. Figure \ref{fig:sdddata} presents these data for selected weeks \citep[for all weeks see Sections 2 and 7 of the Supplemental Material;][]{FullYearOU_Supp}, showing the variable nature of golden eagle migration in western North America, with some of the population migrating south in fall from Alaska and northwestern Canada to far southwestern Canada and the contiguous western U.S. then returning north in spring, some remaining year-round in southern regions, and some inhabiting mid-latitude regions in summer and exhibiting relatively short- to moderate-distance migrations to and from winter ranges.

\begin{figure}[tbp]
    \centering
    \includegraphics[width = 5.61893in]{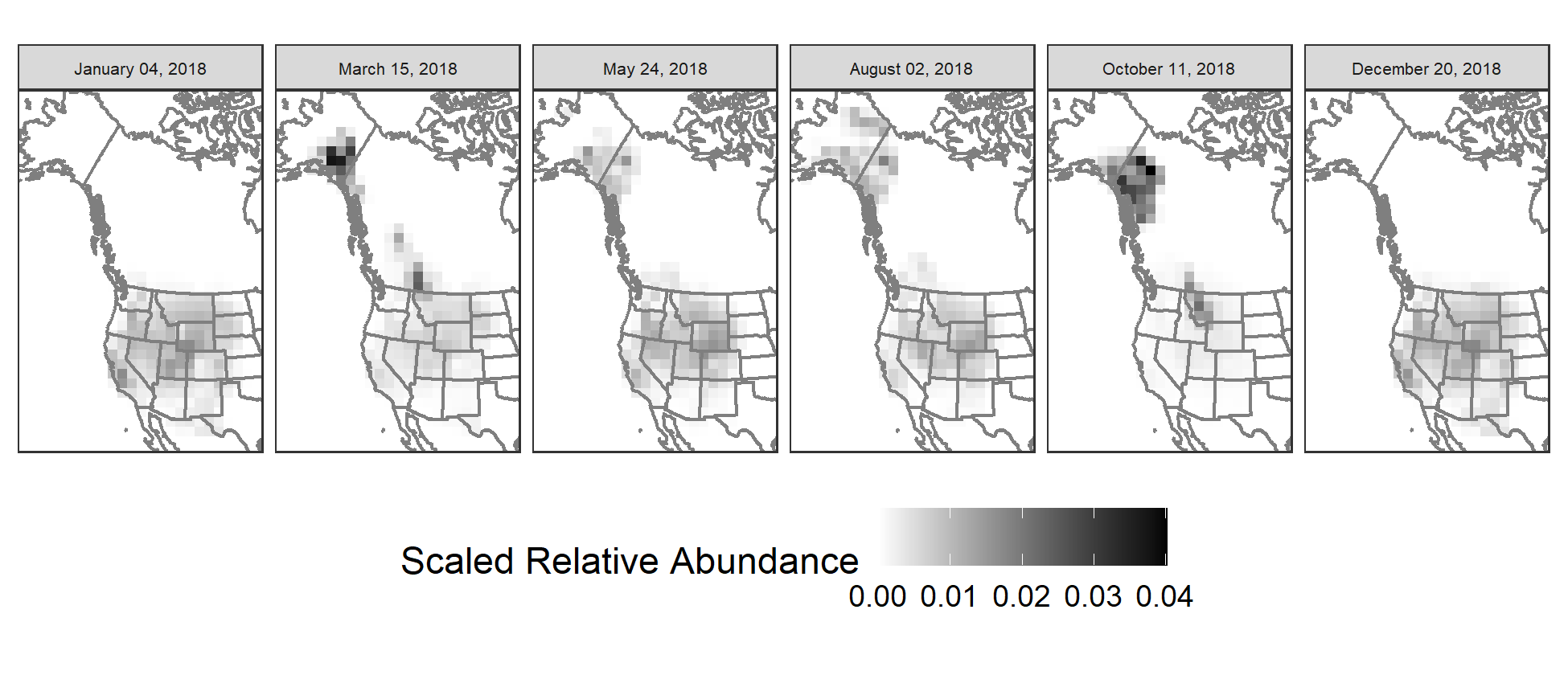}
    \caption{Selected weeks from the 2018 eBird adaSTEM output for golden eagles in western North America, which is used as our species distribution data. This highlights the range of migratory behaviors exhibited by individual golden eagles in the population.}
    \label{fig:sdddata}
\end{figure}

\begin{figure}[tbp]
    \centering
    \includegraphics[width = 5.61893in]{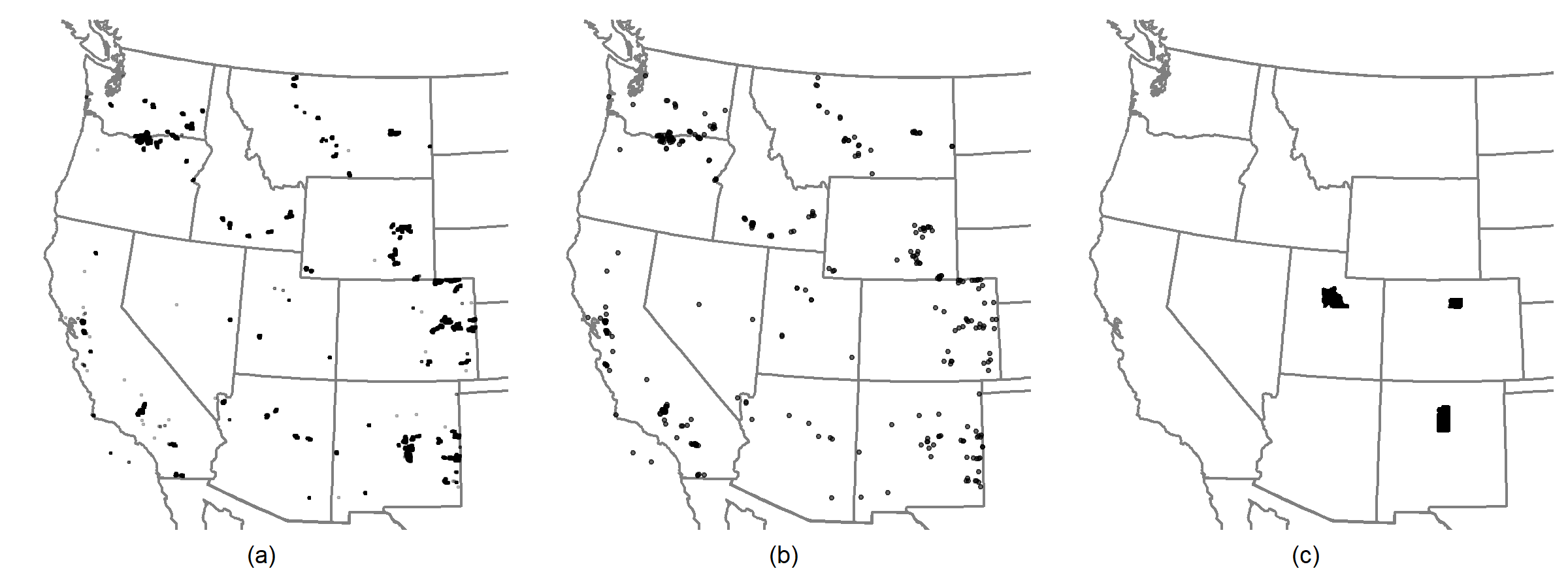}
    \caption{(a). Locations of the 18,456 wind turbines within the range of our species distribution data that are within the contiguous U.S. (b). Locations of the 396 wind projects within the range of our species distribution data that are within the contiguous U.S. (c). Locations of Utah County - Utah (top left), Boulder County - Colorado (top right), and Santa Fe County - New Mexico (bottom).}
    \label{fig:windmills}
\end{figure}

We obtained locations of wind turbines in the U.S. from the U.S. Wind Turbine Database, provided by the U.S. Geological Survey, American Clean Power Association, and Lawrence Berkeley National Laboratory via \url{https://eerscmap.usgs.gov/uswtdb} \citep{windmills}. These services provided locations of commercial wind turbines in the U.S., including characteristics of each such as year when first operational and size (in megawatts). We subset these data to the area of our species distribution data that is within the contiguous U.S., resulting in 18,456 wind turbines that are shown in Figure \ref{fig:windmills}(a). Most turbines are close to one another; because of this, we aggregated turbines to the 396 different wind projects that these 18,456 turbines comprise. A given wind project's location was defined as the centroid of the locations of the wind turbines in that project. These wind project locations are displayed in Figure \ref{fig:windmills}(b); it appears that wind project centroids are a reasonable surrogate for overall turbine distribution. We will evaluate wind project risk for the entire population of eagles as well as specifically for golden eagles that winter in Utah County - Utah, Boulder County - Colorado, and Santa Fe County - New Mexico, which are shown in Figure \ref{fig:windmills}(c).

\section{Methods} \label{sec:methods}
\subsection{Integrated Movement Modeling}
The IMM framework of \citet{IMM_original} suggests one formal approach for scaling individual-level movement models to models for population-level movement. We outline here the general framework we will use for integrated modeling of individual and population data. In Sections \ref{sec:indmovementmodel}--\ref{sec:SDDlikelihood} we will detail the specific models we propose for the eagle data analysis. First consider an individual movement model
\begin{equation}
    \mathbf{x}_{i,y,t+\Delta}\sim f(\mathbf{x}_{i,y,t+\Delta}|\mathbf{x}_{i,y,t},\boldsymbol{\theta}_{i,y},\boldsymbol{\psi}) \label{eq:framework_individual_movement_model}
\end{equation}
which defines the statistical distribution of $\mathbf{x}_{i,y,t+\Delta}$ which is the position of the $i^{\text{th}}$ animal during year $y$ at time $t+\Delta$ measured in days. This model depends on where the animal was at time $t$, individual year-specific parameters $\boldsymbol{\theta}_{i,y}$, and population parameters $\boldsymbol{\psi}$. We then specify a model, $g$, for how the parameters $\boldsymbol{\theta}_{i,y}$ that govern individual movement vary across the population, given population parameters $\boldsymbol{\psi}$
\begin{equation}
    \boldsymbol{\theta}_{i,y}\sim g(\boldsymbol{\theta}_{i,y}|\boldsymbol{\psi}). \label{eq:framework_movement_heterogeneity}
\end{equation}
To obtain a model for population-level movement, we specify an initial spatial distribution $\pi_0(\mathbf{x})$, where $\pi_t(\mathbf{x})$ is the spatial distribution of the population at time $t$. We can then find $\pi_t(\mathbf{x})$ by integrating the individual movement model over the model for individual heterogeneity and the initial distribution
\begin{equation}
    \pi_t(\mathbf{x}) = \int\int f(\mathbf{x}|\mathbf{z},\boldsymbol{\theta}_{i,y},\boldsymbol{\psi})g(\boldsymbol{\theta}_{i,y}|\boldsymbol{\psi})\pi_0(\mathbf{z})d\boldsymbol{\theta}_{i,y}d\mathbf{z} .\label{eq:framework_integral}
\end{equation}
This double integral formulation provides us with a method for scaling up any model for individual tracking data to a movement model for population level data. Finally we specify a model, $h$, for species distribution data which has a formal link to the model used for individual tracking data
\begin{equation}
    s_t(\mathbf{x}) \sim h(s_t(\mathbf{x})|\pi_t(\mathbf{x}),\boldsymbol{\kappa}), \label{eq:framework_sddatamodel}
\end{equation}
where $s_t(\mathbf{x})$ is the eBird adaSTEM output for the grid cell centered at location $\mathbf{x}$ at time $t$ ($t = 4, 11, 18, \ldots 361$ days), which has been normalized to sum to unity for each $t$ as described in Section \ref{sec:data}, and $\boldsymbol{\kappa}$ is a vector of error process parameters. The likelihoods in \eqref{eq:framework_individual_movement_model} and \eqref{eq:framework_sddatamodel} define a joint model for telemetry and species distribution data.
\subsection{Modeling Heterogeneity in Movement} \label{sec:heterogeneity}
Before presenting our time-varying OU process model for individual movement and scaling it up to a population-level movement model, it will be helpful to describe how we will model heterogeneity in movement behavior. We propose modeling individual movement parameters using a mixture model prior, replacing \eqref{eq:framework_movement_heterogeneity} with 
\begin{equation}
    g(\boldsymbol{\theta}_{i,y}|\boldsymbol{\psi}) = \sum_{p=1}^{p}\omega_{ip}g_p(\boldsymbol{\theta}_{i,y}|\boldsymbol{\psi})
\end{equation}
where $\omega_{i1},\ldots,\omega_{iP}$ are nonnegative and sum to unity for each $i$. With the additional assumptions that we can partition $\boldsymbol{\psi}$ into $\boldsymbol{\psi}_1,\ldots, \boldsymbol{\psi}_P$ such that $g_p(\boldsymbol{\theta}_{i,y}|\boldsymbol{\psi})=g_p(\boldsymbol{\theta}_{i,y}|\boldsymbol{\psi}_p)$ and that for each $i$ one of the $\omega_{ip}=1$ and the rest are zero, this mixture model of variation in movement behavior is equivalent to having $P$ subpopulations, each with their own parameters $\boldsymbol{\psi}_p$. We will define our model for movement by subpopulation denoted by a superscript $p$ (or $p_{iy}$ for the subpopulation of individual $i$ in year $y$ when there exists ambiguity). We use $P=4$ subpopulations as described in Section \ref{sec:data}.
\subsection{A Time-Varying OU Model For Full-Annual Cycle Individual Migratory Movement}
\label{sec:indmovementmodel}
We propose a time-varying OU process to model individual animal movement. OU processes have been heavily used for modeling animal movement \citep{dunn_1977_radiotelemetry} as they can capture central place foraging behavior, and other common movement patterns. \citet{blackwell_1997} introduced a time-varying extension by considering a finite-state continuous-time Markov chain where each state corresponds to a different movement pattern (e.g. an OU process) and \citet{Blackwell_2003_comp} provides the tools to perform inference for such models. \citet{IMM_original} used an OU process for modeling spring migration of golden eagles, but only considered eight weeks of the year, used a time-homogeneous SDE, and used a numerical approximation to the SDE. In this work, we develop a time-varying OU process model for the full annual cycle of migratory species which has a fully analytic solution that enables efficient computing. Since the risks (wind projects, weather, etc.) that migratory species encounter vary throughout the year, principled management requires an understanding of how individuals move over the course of the year. As the general timing of migration is well studied for many migratory species, we propose an OU process model with a time-varying attractive point which controls when and where birds migrate as follows. 

Let $\mathbf{x}_{i,y,t}^{p}$ denote the telemetry location at time $t$ (in days) for bird $i$ during year $y$ in subpopulation $p$. We then propose a model for individual movement using the SDE
\begin{equation}
\label{eq:SDE}
    d\mathbf{x}_{i,y,t}^{p} = \begin{cases} -\theta^{p} (\mathbf{x}_{i,y,t}^{p} - \mathbf{m}_{iy1}^{p})dt + \sigma^{p}d\mathbf{W}_{iyt}^{p} & \text{if } t < b_{iy1}^{p} \\
    -\theta^{p} (\mathbf{x}_{i,y,t}^{p} - \mathbf{m}_{iy2}^{p})dt + \sigma^{p}d\mathbf{W}_{iyt}^{p} & \text{if } b_{iy1}^{p} < t < b_{iy2}^{p} \\
    -\theta^{p} (\mathbf{x}_{i,y,t}^{p} - \mathbf{m}_{iy3}^{p})dt + \sigma^{p}d\mathbf{W}_{iyt}^{p} & \text{if } t > b_{iy2}^{p}
    \end{cases}.
\end{equation}
Here the drift term is the gradient of a quadratic potential function that is centered at bird-specific winter attractive points $\mathbf{m}_{iy1}^{p}$ and $\mathbf{m}_{iy3}^{p}$ and summer attractive point $\mathbf{m}_{iy2}^{p}$ depending on the time of year \citep{preisler_2013, Russell_2018, eisenhauer_2022}. The bird-specific migration timing parameters $b_{iy1}^{p}$ and $b_{iy2}^{p}$ correspond to when the bird starts migrating in the spring and fall, respectively. Under this model, mean movement is in the direction of the attractive points with random variation modeled through 2-dimensional Brownian motion ($\mathbf{W}_{iyt}^{p}$). This model can be seen as a special case of the model of \citet{blackwell_1997} with three states each corresponding to an OU process with different attractive points and transition times of $b_{iy1}^{p}$ and $b_{iy2}^{p}$. Each of these three OU processes can also be seen as overdamped Langevin diffusion with speed parameter $\sigma^p$ and a bivariate Gaussian stationary distribution with mean equal to the appropriate attractive point and variance $\frac{(\sigma^p)^2}{2\theta^p}\mathbf{I}$ \citep{michelot2019langevin}.  This model is a time-varying OU process which has an exact solution \citep{gardiner_2009}, depending on the time interval and migration timing parameters:
\begin{align}
    \label{eq:OUcondSolution}
    \mathbf{x}_{i,y,t+\Delta}^{p}|\mathbf{x}_{i,y,t}^{p} \sim \mathcal{N}\biggl(&e^{-\theta^{p}\Delta}\mathbf{x}_{i,y,t}^{p}+\gamma_{iyt\Delta1}^p\mathbf{m}_{iy1}^{p}+\gamma_{iyt\Delta2}^p\mathbf{m}_{iy2}^{p} + \gamma_{iyt\Delta3}^p\mathbf{m}_{iy3}^{p},  \\
    &\frac{(\sigma^{p})^2}{2\theta^{p}}(1-e^{-2\theta^{p}\Delta})\mathbf{I}\biggr) \nonumber
\end{align}
for $\Delta\leq365-t$, where 
\begin{equation}
\gamma_{iyt\Delta3}^p = 1-e^{-\theta^{p}\Delta}-\gamma_{iyt\Delta1}^p - \gamma_{iyt\Delta2}^p,
\end{equation}
\begin{align}
    \gamma_{iyt\Delta1}^p &= \begin{cases}
        1 - e^{-\theta^{p}\Delta} & t < t+\Delta \leq b^{p}_{iy1} < b^{p}_{iy2} \\
        (1 - e^{-\theta^{p}(b^{p}_{iy1}-t)})e^{-\theta^{p}(t+\Delta-b^{p}_{iy1})} & t < b^{p}_{iy1} < t+\Delta < b^{p}_{iy2} \\
        0 &  b^{p}_{iy1} \leq t < t+\Delta \leq b^{p}_{iy2} \\
        0 &  b^{p}_{iy1} < t < b^{p}_{iy2} < t+\Delta \\
        0 &  b^{p}_{iy1} < b^{p}_{iy2} \leq t < t+\Delta \\
        (1-e^{-\theta^{p}(b_{iy1}^{p}-t)})e^{-\theta^{p}(t+\Delta-b_{iy1}^{p})} &  t<b_{iy1}^{p}<b_{iy2}^{p}<t+\Delta
    \end{cases}, \label{eq:witdelta}
\end{align}
and 
\begin{align}
    \gamma_{iyt\Delta2}^p &= \begin{cases}
        0 & t < t+\Delta \leq b^{p}_{iy1} < b^{p}_{iy2} \\
        1 - e^{-\theta^{p}(t+\Delta-b^{p}_{iy1})} & t < b^{p}_{iy1} < t+\Delta < b^{p}_{iy2} \\
        1 - e^{-\theta^{p}\Delta} &  b^{p}_{iy1} \leq t < t+\Delta \leq b^{p}_{iy2} \\
        (1 - e^{-\theta^{p}(b^{p}_{iy2}-t)})e^{-\theta^{p}(t+\Delta-b^{p}_{iy2})} &  b^{p}_{iy1} < t < b^{p}_{iy2} < t+\Delta \\
        0 &  b^{p}_{iy1} < b^{p}_{iy2} \leq t < t+\Delta \\
        (1 - e^{-\theta^{p}(b^{p}_{iy2}-b^{p}_{iy1})})e^{-\theta^{p}(t+\Delta-b^{p}_{iy2})} &  t<b_{iy1}^{p}<b_{iy2}^{p}<t+\Delta
    \end{cases}. \label{eq:sitdelta}
\end{align}
In this model, the conditional mean of a bird's location at time $t+\Delta$ given where it was at time $t$ is a weighted average of its location at time $t$ $(\mathbf{x}_{i,t})$, its winter attractive points ($\mathbf{m}_{iy1}^{p}$ and $\mathbf{m}_{iy3}^{p}$), and its summer attractive point ($\mathbf{m}_{iy2}^{p}$), where $\gamma_{iyt\Delta1}^p$, $\gamma_{iyt\Delta2}^p$, and $\gamma_{iyt\Delta3}^p$ control the weights on the respective winter and summer attractive points. The six different cases in \eqref{eq:witdelta} and \eqref{eq:sitdelta} characterize the behavior of the bird during the time window $(t,t+\Delta)$. In the first case, the interval $(t,t+\Delta)$ is completely before the bird starts its spring migration, and so the bird is attracted only to its first winter attractive point, $\mathbf{m}_{iy1}^{p}$, $\gamma_{iyt\Delta1}^p = 1-e^{-\theta^{p}\Delta}$, $\gamma_{iyt\Delta2}^p=0$, and $\gamma_{iyt\Delta3}^p=0$. This is the classic result for time homogeneous OU processes \citep[e.g.][]{gardiner_2009}. In the second case, the interval $(t,t+\Delta)$ contains $b_{iy1}^{p}$, when the bird starts its spring migration, and the movement is a composite of homogeneous movement with mean movement toward $\mathbf{m}_{iy1}^{p}$ for the interval $(t,b_{iy1}^{p})$ followed by homogeneous movement with mean movement toward $\mathbf{m}_{iy2}^{p}$ for the interval $(b_{iy1}^{p},t+\Delta)$. The remaining four cases are similar, with behavior being a composition of homogeneous movement between different breakpoints where behavior shifts with the start of spring and/or fall migrations. This model gives us a likelihood for each telemetry data point conditional on the most recent telemetry observation. For the initial location of bird $i$ in subpopulation $p$, we assume a priori that
\begin{equation}
    x_{i0}^{p}\sim\mathcal{N}\left(\mathbf{m}_{iy1}^{p},\frac{(\sigma^{p})^2}{2\theta^{p}}\mathbf{I}\right) \label{eq:ind_init}
\end{equation}
which is the stationary distribution of the model described in \eqref{eq:OUcondSolution} before the spring migration when the OU process has mean movement toward $\mathbf{m}_{iy1}^{p}$ \citep[or equivalently the stationary distribution of the Langevin diffusion that makes up the first case of the SDE in \eqref{eq:SDE};][]{michelot2019langevin}. The individual movement model in \eqref{eq:OUcondSolution}-\eqref{eq:ind_init} captures the full annual cycle of individual migratory behavior using an analytically tractable time-varying OU model and is, to our knowledge, the first SDE model of full-annual cycle behavior in migratory animals.
\subsection{Modeling Heterogeneity in Movement Behavior}
We now specify our model for heterogeneity in movement behavior of individuals within each subpopulation. As described in Section \ref{sec:heterogeneity}, each subpopulation can also equivalently be thought of as a component of our mixture model for variation in movement behavior. Our GPS data consist of 215 bird-years from 93 unique eagles.  We assume that each individual's winter and summer attractive points are independent of all other individuals. When one eagle is observed over multiple seasons (including winter to winter in a single calendar year), we propose modeling each season's attractive points with time series dependence to capture possible site fidelity \citep[e.g.][]{SiteFidelity}. Thus, we propose to model the winter ($\mathbf{m}_{iy1}^p$ and $\mathbf{m}_{iy3}^p$) and summer ($\mathbf{m}_{iy2}^p$) attractive points using OU processes
\begin{align}
    d\mathbf{m}_{iw}^p &= -\delta_w(\mathbf{m}_{iw}^p-\boldsymbol{\mu}_w^p)dt + \sqrt{2\delta_w}(\boldsymbol{\Sigma}_w^p)^{1/2}d\mathbf{W}_t \\
    d\mathbf{m}_{is}^p &= -\delta_s(\mathbf{m}_{is}^p-\boldsymbol{\mu}_s^p)dt + \sqrt{2\delta_s}(\boldsymbol{\Sigma}_s^p)^{1/2}d\mathbf{W}_t.
\end{align}
These OU processes have stationary distributions which we use as the prior distribution for each individual bird's first winter and summer attractive points; therefore, if $y$ is the first year for which we have data for bird $i$,
\begin{align}
    \mathbf{m}_{iy1}^{p} &\sim \mathcal{N}\left(\boldsymbol{\mu}_w^{p},\boldsymbol{\Sigma}_w^{p}\right) & 
    \mathbf{m}_{iy2}^{p} &\sim \mathcal{N}\left(\boldsymbol{\mu}_s^{p},\boldsymbol{\Sigma}_s^{p}\right).
    \label{eq:attractivenessprior_origyear}
\end{align}
If we have the data for year $y-1$ for bird $i$, then the first winter and summer attractive points are distributed 
\begin{align}
    \mathbf{m}_{iy1}^{p} &= \mathbf{m}_{i(y-1)3}^p  \label{eq:attractivenessprior_haveprevious_1} \\
    \mathbf{m}_{iy2}^{p} &\sim \mathcal{N}\left(e^{-\delta_s}\mathbf{m}_{i(y-1)2}^{p} + (1-e^{-\delta_s}) \boldsymbol{\mu}_s^{p_{iy}},(1-e^{-2\delta_s})\boldsymbol{\Sigma}_s^{p_{iy}}\right).
    \label{eq:attractivenessprior_haveprevious_2}
\end{align}
Then if $y$ is not the first year for which we have data for bird $i$ but we do not have data for year $y-1$ the first winter and summer attractive points are distributed
\begin{align}
    \mathbf{m}_{iy1}^{p} &\sim \mathcal{N}\left(e^{-\delta_w(y-y'-1)}\mathbf{m}_{iy'3}^{p} + (1-e^{-\delta_w(y-y'-1)}) \boldsymbol{\mu}_w^{p_{iy}},(1-e^{-2\delta_w(y-y'-1)})\boldsymbol{\Sigma}_w^{p_{iy}}\right) \label{eq:attractivenessprior_other_1} \\
    \mathbf{m}_{iy2}^{p} &\sim \mathcal{N}\left(e^{-\delta_s(y-y')}\mathbf{m}_{iy'2}^{p} + (1-e^{-\delta_s(y-y')}) \boldsymbol{\mu}_s^{p_{iy}},(1-e^{-2\delta_s(y-y')})\boldsymbol{\Sigma}_s^{p_{iy}}\right),
    \label{eq:attractivenessprior_other_2}
\end{align}
where $y'$ is the most recent year for which we have data for bird $i$. Finally the second winter attractive point has distribution
\begin{equation}
    \mathbf{m}_{iy3}^{p} \sim \mathcal{N}\left(e^{-\delta_w}\mathbf{m}_{iy1}^{p} + (1-e^{-\delta_w}) \boldsymbol{\mu}_w^{p},(1-e^{-2\delta_w})\boldsymbol{\Sigma}_w^{p}\right). \label{eq:attractivenessprior_3}
\end{equation}
Combining \eqref{eq:OUcondSolution}, \eqref{eq:attractivenessprior_origyear}, and \eqref{eq:attractivenessprior_3} with standard multivariate normal theory allows us to marginalize over the individual attractive points and our marginalized SDE model for individual movement for an individual during its first year of observed data is
\begin{align}
 \label{eq:OUintegrated}
     \mathbf{x}_{i,y,t+\Delta}^{p}|\mathbf{x}_{i,y,t}^{p} \sim \mathcal{N}\biggl(&\mathbf{x}_{i,y,t}^{p}e^{-\theta^{p}\Delta}+(\gamma_{iyt\Delta1}^p+\gamma_{iyt\Delta3}^p)\boldsymbol{\mu}_w^{p}+\gamma_{iyt\Delta2}^p\boldsymbol{\mu}_s^{p}, \\
     &\frac{(\sigma^{p})^2}{2\theta^{p}}(1-e^{-2\theta^{p}\Delta})\mathbf{I} +(\gamma_{iyt\Delta2}^p)^2\boldsymbol{\Sigma}_s^{p} + \nonumber \\
     & \Bigl((\gamma_{iyt\Delta1}^p)^2 + (\gamma_{iyt\Delta3}^p)^2 + 2\gamma_{iyt\Delta1}^p\gamma_{iyt\Delta3}^pe^{-\delta_w}\Bigr)\boldsymbol{\Sigma}_w^{p} \biggr) . \nonumber
 \end{align}
In our inference below, we use the model described in \eqref{eq:OUcondSolution} rather than \eqref{eq:OUintegrated} as our likelihood for the telemetry data. However, we present \eqref{eq:OUintegrated} as it is useful for motivating the initial species distribution of Section \ref{sec:InitialSpecies} and is an intermediate step for the resulting population-level movement model of Section \ref{sec:resultingPopDyn}. 

We now specify our prior for the individual migration timings. We allow each subpopulation to begin their spring migration between January 15 (Julian day 15) and June 15 (Julian Day 166) and their fall migration between July 15 (Julian day 196) and December 15 (Julian Day 350) \citep[expanded from][]{bedrosian_2018}. We then allow each bird within each subpopulation to vary when they migrate, with migration timing following a scaled and shifted beta distribution with subpopulation specific shape parameters.
\begin{align}
    b_{iy1}^{p} &= 151a_{i1}^{p} + 15 & b_{iy2}^{p} &= 154a_{i2}^{p} + 196 \label{eq:timingstuff} \\
    a_{i1}^{p} &\sim \text{Beta}(\alpha_1^{p},\alpha_2^{p}) & a_{i2}^{p} &\sim \text{Beta}(\alpha_3^{p},\alpha_4^{p}) \label{eq:timingdist}
\end{align}
Thus we have a hierarchal mixture-model style prior for our individual-specific parameters. This model allows for each individual bird to have different migratory behavior while still allowing us to scale up our individual SDE model to an SDE model for population level movement. Modeling heterogeneity in movement behavior with the mixture model prior (with each mixture component being referred to as a subpopulation) results in a very flexible model for variation in behavior across the population, with interpretable parameters that can be modeled using priors derived from known information about the species (i.e., the bounds on migration timing parameters).

\subsection{Initial Species Distribution} \label{sec:InitialSpecies}
In the integrated movement modeling framework we propose here (Section \ref{sec:resultingPopDyn}), we model the spatio-temporal population dynamics by marginalizing the joint distribution of how all individuals in the population move, \eqref{eq:OUcondSolution} and \eqref{eq:ind_init}, over the variation in population behavior. To do this, we need to specify an initial spatial distribution for the species distribution. We propose a Gaussian mixture model, with 
\begin{equation}
    \pi_0(\mathbf{x}) = \sum_{p=1}^4\eta^{p}\pi_0^{p}(\mathbf{x})
    \label{eq:totalinit}
\end{equation}
where $\boldsymbol{\eta}=\begin{bmatrix}\eta^1&\eta^2&\eta^3&\eta^4\end{bmatrix}^T$ is a vector of nonnegative weights that sum to unity and  $\pi_0^{p}(\mathbf{x})$ is the initial distribution of the birds in subpopulation $p$. We model $\pi_0^{p}(\mathbf{x})$ as a Gaussian distribution based off of the stationary distribution of the SDE model defined in \eqref{eq:OUintegrated}
\begin{equation}
    \pi_0^{p}(\mathbf{x}) = \Phi\Bigl(\mathbf{x};\boldsymbol{\mu}_w^{p},\frac{(\sigma^{p})^2}{2\theta^{p}}\mathbf{I} + \boldsymbol{\Sigma}_w^{p}\Bigr) .
    \label{eq:InitialDynamics}
\end{equation}
where $\Phi(\mathbf{y};\boldsymbol{\nu},\mathbf{V})$ is the density of a bivariate Gaussian distribution with mean $\boldsymbol{\nu}$ and variance-covariance matrix $\mathbf{V}$ evaluated at $\mathbf{y}$. Thus, our model for the initial (Jan. 1) distribution of the population distribution is a Gaussian mixture model with each component being the stationary distribution of the individual movement model before migration, marginalizing over variation in subpopulation attractive points.

\subsection{Resulting Population-level Movement Model} \label{sec:resultingPopDyn}
By integrating over the model for variation in movement behavior, \eqref{eq:attractivenessprior_origyear}, \eqref{eq:attractivenessprior_3}, and \eqref{eq:timingdist}, and initial spatial distribution for the species distribution \eqref{eq:totalinit} we can scale up our above individual movement model \eqref{eq:OUcondSolution} to a model for the spatio-temporal dynamics of the species distribution. Due to the Gaussian structure in our initial spatial distribution and our model for movement heterogeneity, we can perform this integration for each mixture component then take a weighted sum of the resulting dynamics. Following \eqref{eq:framework_integral} we propose calculating for each subpopulation
\begin{align}
    \pi_t^{p}(\mathbf{x}) = \int\int\int\int\int &\Bigl[f(\mathbf{x}_{i,y,t}^{p}|\mathbf{x}_{i,0}^{p}=\mathbf{z})g(\mathbf{m}_{iy1}^{p}|\boldsymbol{\mu}_w^{p},\boldsymbol{\Sigma}_w^{p})g(\mathbf{m}_{iy2}^{p}|\boldsymbol{\mu}_s^{p},\boldsymbol{\Sigma}_s^{p})g(\mathbf{m}_{iy3}^{p}|\boldsymbol{\mu}_w^{p},\boldsymbol{\Sigma}_w^{p})   \nonumber  \\ 
    &g(b_{iy1}^{p}|\alpha_1^{p},\alpha_2^{p})g(b_{iy2}^{p}|\alpha_3^{p},\alpha_4^{p}) \pi_0^{p}(\mathbf{z})\Bigr]d\mathbf{m}_{iy1}^{p}d\mathbf{m}_{iy2}^{p}db_{iy1}^{p}db_{iy2}^{p}d\mathbf{z}, \label{eq:5int}
\end{align}
which is the marginal spatial distribution of mixture component $p$ at time $t$, marginalizing over the variation in individual movement behavior and the initial spatial distribution of that mixture component. As $f(\mathbf{x}_{i,y,t}^{p}|\mathbf{x}_{i,0}^{p}=\mathbf{z})$, $g(\mathbf{m}_{iy1}^{p}|\boldsymbol{\mu}_w^{p},\boldsymbol{\Sigma}_w^{p})$, $g(\mathbf{m}_{iy2}^{p}|\boldsymbol{\mu}_s^{p},\boldsymbol{\Sigma}_s^{p})$,
$g(\mathbf{m}_{iy3}^{p}|\boldsymbol{\mu}_w^{p},\boldsymbol{\Sigma}_w^{p})$, and $\pi_0^{p}(z)$ are all bivariate Gaussian distributions, if we assume that the migration timings ($b_{iy1}^{p}$ and $b_{iy2}^{p}$) are known, combining \eqref{eq:OUintegrated} and \eqref{eq:InitialDynamics} with standard multivariate Gaussian distribution theory results in an analytic form for the spatial distribution for mixture component $p$ at time $t$, 
\begin{align}
    \pi_t^{p}(\mathbf{x}|b_{iy1}^{p},b_{iy2}^{p}) &= \int\int\int \Bigl[f(\mathbf{x}_{i,y,t}^{p}|\mathbf{x}_{i,0}^{p}=\mathbf{z})g(\mathbf{m}_{iy1}^{p}|\boldsymbol{\mu}_w^{p},\boldsymbol{\Sigma}_w^{p})g(\mathbf{m}_{iy2}^{p}|\boldsymbol{\mu}_s^{p},\boldsymbol{\Sigma}_s^{p}) \label{eq:3int} \\
    & \hspace*{.75in} g(\mathbf{m}_{iy3}^{p}|\boldsymbol{\mu}_w^{p},\boldsymbol{\Sigma}_w^{p}) \pi_0^{p}(\mathbf{z})\Bigr]d\mathbf{m}_{iy1}^{p}d\mathbf{m}_{iy2}^{p}d\mathbf{z} \label{eq:3int} \nonumber \\
    &=\Phi\biggl(\mathbf{x};\bigl(e^{-\theta^{p}t}+\gamma_{iy0t1}^p+\gamma_{iy0t3}^p\bigr)\boldsymbol{\mu}_w^{p}+\gamma_{iy0t2}^p\boldsymbol{\mu}_s^{p}, \nonumber \\ 
    &\hspace*{.6in}\frac{(\sigma^{p})^2}{2\theta^{p}}\mathbf{I} +(\gamma_{iy0t2}^p)^2\boldsymbol{\Sigma}_s^{p} + \nonumber \\
     &\hspace*{.6in} \Bigl(e^{-2\theta t} + (\gamma_{iy0t1}^p)^2 + (\gamma_{iy0t3}^p)^2 + 2\gamma_{iy0t1}^p\gamma_{iy0t3}^pe^{-\delta_w}\Bigr)\boldsymbol{\Sigma}_w^{p}\biggr). \nonumber
\end{align}
Thus conditional on migration timing, we have an analytic Gaussian form for the spatio-temporal dynamics of $P$ subpopulations of animals that each move following our time-varying SDE model for full annual cycle migratory movement with mixture model variation in individual parameters. 

We cannot marginalize over the individual migration timings analytically in the same manner that we did with the individual migration attractive points. Instead, we propose to use quasi Monte-Carlo techniques \citep{morokoff_1995_quasi} to approximate the integral marginalizing over the distributions of spring and fall migration timings:
\begin{equation}
    \pi_t^{p}(\mathbf{x}) = \int\int \pi_t^{p}(\mathbf{x}|b_{iy1}^{p},b_{iy2}^{p})g(b_{iy1}^{p}|\alpha_1^{p},\alpha_2^{p})g(b_{iy2}^{p}|\alpha_3^{p},\alpha_4^{p})db_{iy1}^{p}db^{p}_{iy2}. 
\end{equation}
We denote our proposed quasi Monte-Carlo approximation as $\Tilde{\pi}_t^{p}(\mathbf{x})$. To calculate this approximation we draw $(a_{11}^{p*}, a_{12}^{p*})$, $(a_{21}^{p*},a_{22}^{p*})$, $\ldots$, $(a_{Q1}^{p*},a_{Q2}^{p*})$ from the Halton sequence \citep{halton_1964} with base 2 for the first component and base 3 for the second component and calculate corresponding $(b_{11}^{p*},b_{12}^{p*})$, $(b_{21}^{p*},b_{22}^{p*})$, $\ldots$, $(b_{Q1}^{p*},b_{Q2}^{p*})$ by following \eqref{eq:timingstuff}. We use $Q=10$. By using a low-discrepancy sequence, such as the Halton sequence, we consistently have a sample of design points that is dispersed on $[0,1]\times[0,1]$ to capture the distribution systematically, with increasing precision as the number of points grows. Then we approximate the double integral with a quasi Monte-Carlo estimate based on our points drawn from the Halton sequence,
\begin{equation}
    \Tilde{\pi}_t^{p}(\mathbf{x}) = \sum_{q=1}^Q\pi_t^{p}(\mathbf{x}|b_{q1}^{p*},b_{q2}^{p*})\frac{g(b_{q1}^{p*}|\alpha_1^{p},\alpha_2^{p})g(b_{q2}^{p*}|\alpha_3^{p},\alpha_4^{p})}{\sum_{j=1}^Qg(b_{j1}^{p*}|\alpha_1^{p},\alpha_2^{p})g(b_{j2}^{p*}|\alpha_3^{p},\alpha_4^{p})},
\end{equation}
where $g(b_{q1}^{p*}|\alpha_1^{p},\alpha_2^{p})$ and $g(b_{q2}^{p*}|\alpha_3^{p},\alpha_4^{p})$ are the beta likelihoods of \eqref{eq:timingdist}. We can then approximate the spatial distribution for the entire population of golden eagles at time $t$ by taking the weighted sum of the species distributions approximations for each subpopulation:
\begin{equation}
    \pi_t(\mathbf{x}) = \sum_{p}\eta^{p}\Tilde{\pi}_t^{p}(\mathbf{x}). 
    \label{eq:speciesdistribution}
\end{equation}
This allows us to estimate the population-level movement, which will be necessary to evaluate the species distribution likelihood, simply by evaluating bivariate Gaussian densities without needing to numerically solve any differential equations. Thus conducting inference which requires repeatedly evaluating the likelihood for the species distribution data \eqref{eq:sdd_ll_final} will be much faster than if we did not have an analytic solution and had to numerically approximate the solution to the SDE.

\subsection{Species Distribution Likelihood}
\label{sec:SDDlikelihood}
Having developed a model for the spatio-temporal dynamics of the population of a migratory species, we now propose a likelihood for the observed species distribution data. Let $s_t(\mathbf{x})$ be the eBird adaSTEM output for the grid cell centered at location $\mathbf{x}$ at time $t$ ($t = 4, 11, 18, \ldots 361$ days), which has been normalized to sum to unity for each $t$ as described in Section \ref{sec:data}. \citet{IMM_original} suggested using a Poisson likelihood, with
\begin{equation}
    \Tilde{s}_t(\mathbf{x}) \sim \text{Poisson}(K\times\pi_t(\mathbf{x})), \label{eq:poisson_bad}
\end{equation}
where $\Tilde{s}_t(\mathbf{x}) = K\times s_t(\mathbf{x})$, rounded to the nearest whole number. We found that the above likelihood for observed relative abundance can be unidentifiable in practice for some datasets. The issue arises from the rounding of $K\times s_t(\mathbf{x})$. When $K$ becomes very small, the scaled data becomes zero for all (or almost all) cells, and the mean of the Poisson likelihood in \eqref{eq:poisson_bad} also becomes very small, making zero observations very likely. Thus, the above likelihood can be problematic, as it can be maximized in the limit as $K\to0$, where the likelihood of each grid cell is 1. 

Instead we propose a new likelihood for the observed eBird adaSTEM output, $s_t(\mathbf{x})$. We assume that the eBird adaSTEM output at each grid cell is an unbiased estimate of the true relative abundance of golden eagles in the study area, and propose
\begin{equation}
    s_t(\mathbf{x}) \sim \mathcal{N}\left(\frac{\pi_t(\mathbf{x})}{\sum_{\mathbf{y}\in\mathcal{D}}\pi_t(\mathbf{y})},\kappa^2\right), \label{eq:sdd_ll_final}
\end{equation}
where $\mathcal{D}$ is the set of centers of all of the grid cells in the adaSTEM output and $\kappa$ is a variance parameter. This new likelihood is easy to implement, has clear assumptions about the data collection process, and is free of the identifiability problems present in the scaled Poisson likelihood in \eqref{eq:poisson_bad}.
    
\subsection{Bayesian Inference}
In the above sections, we have specified a joint model for individual telemetry data and spatio-temporal species distribution data, where the model consists of a time-varying SDE model for individual movement, with individual movement parameters modeled using a mixture model. We now propose an approach for Bayesian inference on the individual and population level parameters in this model. The model that we fit is composed of \eqref{eq:OUcondSolution}, \eqref{eq:ind_init}, \eqref{eq:attractivenessprior_origyear}-\eqref{eq:attractivenessprior_3},\eqref{eq:timingdist}, and \eqref{eq:sdd_ll_final}. For fitting the model we propose a Bayesian approach where we take draws from the posterior distribution using Markov chain Monte Carlo. To fully specify our model we must specify priors for all model parameters: we do that here. For the subpopulation attractiveness parameter we use a transformed uniform distribution
\begin{equation}
    e^{-\theta^p} \stackrel{iid}{\sim}\text{Uniform}(0.02, 0.98).
\end{equation}
We do this to prevent the limit cases where either the previous telemetry location ($\theta^p\to \infty$) or attractive points ($\theta^p\to0$) have no effect on the mean of the telemetry likelihood, \eqref{eq:OUcondSolution}. For the subpopulation variances, site fidelity parameters, and species distribution error variance, which all must be positive, we use diffuse log-normal priors
\begin{align}
    (\sigma^{p})^2&\stackrel{iid}{\sim}\text{log-normal}(0,10^6) \\
    \delta_w&\stackrel{iid}{\sim}\text{log-normal}(0,10^6) \\
    \delta_s&\stackrel{iid}{\sim}\text{log-normal}(0,10^6) \\
    \kappa^2&\stackrel{iid}{\sim}\text{log-normal}(0,10^6).
\end{align}
For $\boldsymbol{\mu}_w^p$ and $\boldsymbol{\mu}_s^p$, the means of the limiting distributions of the OU processes for the winter and summer attractive points, we use a uniform distribution over the area covered by the eBird data. For $\boldsymbol{\Sigma}_w^p$ and $\boldsymbol{\Sigma}_w^p$, the variances of the limiting distributions of the OU processes for the winter and summer attractive points we have
\begin{align}
    \boldsymbol{\Sigma}_w^{p} &= \begin{bmatrix}
        e^{\tau_{w1}^p} & \rho_w^pe^{(\tau_{w1}^p+\tau_{w2}^p)/2} \\
        \rho_w^pe^{(\tau_{w1}^p+\tau_{w2}^p)/2} & e^{\tau_{w2}^p}
    \end{bmatrix} &
    \boldsymbol{\Sigma}_s^{p} &= \begin{bmatrix}
        e^{\tau_{s1}^p} & \rho_s^pe^{(\tau_{s1}^p+\tau_{s2}^p)/2} \\
        \rho_s^pe^{(\tau_{s1}^p+\tau_{s2}^p)/2} & e^{\tau_{s2}^p}
    \end{bmatrix} \\
    p(\tau_{w1}^p) &\propto \mathbf{I}\left\{\tau_{w1}^p \leq 28.7 \right\} & p(\tau_{s1}^p) &\propto \mathbf{I}\left\{\tau_{s1}^p \leq 28.7 \right\} \label{eq:tau1}\\
    p(\tau_{w2}^p) &\propto \mathbf{I}\left\{\tau_{w2}^p \leq 29.5 \right\} & p(\tau_{s2}^p) &\propto \mathbf{I}\left\{\tau_{s2}^p \leq 29.5 \right\} \label{eq:tau2} \\
    \rho_w^p &\stackrel{iid}{\sim} \text{Uniform}(0,1) & \rho_s^p &\stackrel{iid}{\sim} \text{Uniform}(0,1)
\end{align}
The upper bounds of the improper uniform distributions for the $\tau$'s, \eqref{eq:tau1} and \eqref{eq:tau2}, are derived by limiting the associated standard deviation to half the range of the eBird data. For the shape parameters of the beta distribution we propose a diffuse truncated log-normal distribution
\begin{equation}
    \ln(\alpha_.^p) \stackrel{iid}{\sim} \text{Truncated-Normal}(0, 10^6, 0, \infty).
\end{equation}
This ensures that $\alpha\geq1$ so the beta density does not go to $\infty$ as you approach 0 or 1, which we want as the periods of allowed migration are wide enough that we don't want the mass concentrated at the extremes. For the proportion of the species distribution in each subpopulation, we use a constrained Dirichlet distribution,
\begin{align}
    \boldsymbol{\eta} &= 0.05\times\mathbf{1} + 0.8\times\boldsymbol{\eta}^* \\
    \boldsymbol{\eta}^*&\sim\text{Dirichlet}(\mathbf{1}).
\end{align}
This makes it so that each subpopulation accounts for between 5\% and 85\% of the total population. To fit this model we use the Metropolis Hastings algorithm with the log adaptive tuning of \citet{shabywells2010adaptive}. We ran 10 MCMC chains in parallel for 300,000 iterations each which took 2,120 hours of CPU time in total. Every 20,000 iterations each chain made an importance resampling update from among the current values of the ten chains \citep{PMC_Import}. Each chain had the first two thirds discarded as burn-in and was thinned by a factor of 10 resulting in 100,000 total MCMC samples. Convergence was assessed visually as well as with effective sample sizes \citep[using the coda R package,][]{R_package_coda}. In general, ESS was high for population parameters and many individual parameters, but low for parameters for individuals that do not closely follow our proposed SDE model for movement; for additional details and effective sample sizes for all parameters, see Section 3 of the Supplemental Material \citep{FullYearOU_Supp}. The posterior predictive p-value based on the usual chi-squared goodness of fit measure \citep[see section 2.2.2 in][]{Cressie_Wikle_spattemp} was 0.512.

\subsection{Assessing Risk Posed by Wind Projects to Golden Eagles in the Western U.S.}
\label{sec:windmill}
We will use this model to quantify the risk that different wind projects pose to golden eagles in the contiguous western United States; the risk is that of an eagle dying from collision with a turbine \citep{Beston_2016,gedir_2025}, specifically a turbine's spinning blade \citep{hunt_2002}. As we do not have mortality data, we instead focus on estimating the wind project risk, which we calculate as the probability of being within the 2.8 x 2.8 km square centered at the centroid of a wind project, which is correlated with, and could be used as a proxy for, collision risk. Let $P_r(t)$ be the probability that an eagle is within the 2.8 x 2.8 km square centered at the centroid of wind project $r$ at time $t$, which can be found by integrating $\pi_t(\mathbf{x})$ over the relevant spatial region. We use the algorithm of \cite{donnelly_1973} for computation of the bivariate normal cdf \citep[as implemented in the VGAM R package;][]{Rpack_VGAM_1, Rpack_VGAM_2}. This gives us a picture of how wind project risk evolves throughout the calendar year. Let 
\begin{equation}
    Q_r = \int_0^{365} P_r(t)dt
\end{equation}
be the occupancy time \citep{Kulkarni_2011}: the mean number of days in a year that an eagle spends in the 2.8 x 2.8 km square centered at the centroid of wind project $r$. This provides a measure comparable across the different wind projects as to how much risk they pose to Golden Eagles. We could use our fitted model to estimate $Q_r$ and related quantities, but the eBird data themselves could be used to approximate $Q_r$, with 
\begin{equation}
Q_r\approx\frac{365}{52}\sum_{t\in\{4,11,\ldots,361\}}\Tilde{s}_t(\mathbf{x}_r),
\end{equation}
where $\Tilde{s}_t(\mathbf{x}_r)$ is the original eBird adaSTEM output at the 2.8 km x 2.8 km resolution for the grid cell that contains the centroid of wind project $r$ and the summation is taken over the points in time where the weekly abundance is estimated.  Our integrated model results in more than just an estimate of the population mass in space and time, it gives us an estimate of the spatio-temporal dynamics of the entire population. This allows us to answer questions about subsets of the population that are not possible to address using the eBird population data alone. For example, management of many species is done by state agencies tasked with monitoring and maintaining a species in their state. Thus, a state might be interested in understanding which wind projects pose the highest risk to birds who winter in their state. eBird data alone cannot answer this question, but the dynamic model resulting from our IMM approach can. As such we will consider wind project risk for eagles that winter in three counties in the western U.S.: Utah County - Utah, Boulder County - Colorado, and Santa Fe County - New Mexico. These three counties are shown in Figure \ref{fig:windmills}(c). To estimate the spatio-temporal distribution for eagles wintering in a given location we fix the winter attractive points, $\mathbf{m}_{iy1}^{p}$ and $\mathbf{m}_{iy3}^{p}$, for all four subpopulations as a point mass at the centroid of the county in question. We will call this value $\mathbf{m}_w$. We also set the initial spatial distribution of subpopulation $p$, $\pi_0^{p}(\mathbf{x})$, to be a point mass at $\mathbf{m}_w$. Let $\Tilde{\xi}^{p}$ be the density value of the prior of $\mathbf{m}_{iy1}^{p}$, given in \eqref{eq:attractivenessprior_origyear}, evaluated at $\mathbf{m}_w$. Then let
\begin{equation}
    \xi^{p} = \frac{\Tilde{\xi}^{p}}{\sum_{i=1}^4\Tilde{\xi}^{p}}
\end{equation}
be the proportion of eagles at location $\mathbf{m}_w$ who are from subpopulation $p$. We then change our initial species distribution, given in 
\eqref{eq:totalinit}, to
\begin{equation}
    \pi_0(\mathbf{x}) = \sum_{p=1}^4\xi^{p}\pi_0^{p}(\mathbf{x}).
\end{equation}
This reflects an assumption that the proportion of individuals from a given subpopulation at a certain location at the year's beginning is based solely on the distribution of the first winter attractive point. These changes result in \eqref{eq:3int} becoming
\begin{align}
    \pi_t^{p}(\mathbf{x})|b_{iy1}^{p},b_{iy2}^{p} \sim\mathcal{N}\biggl(&(e^{-\theta^{p}t}+\gamma_{iy0t1}+\gamma_{iy0t3})\mathbf{m}_{w} + \gamma_{iy0t2}\boldsymbol{\mu}_s^{p}, \\
    &\frac{(\sigma^{p})^2}{2\theta^{p}}(1-e^{-2\theta^{p}t})\mathbf{I} + (\gamma_{iy0t2})^2\boldsymbol{\Sigma}_s^{p}\biggr) \nonumber
\end{align}
and \eqref{eq:speciesdistribution} becoming
\begin{equation}
    \pi_t(\mathbf{x}) = \sum_{p=1}^4\xi^{p}\Tilde{\pi}_t^{p}(\mathbf{x}).
\end{equation}
All other results from the section \ref{sec:resultingPopDyn} remain unchanged. Let $P_{jr}(t)$ be the probability that an eagle that winters in county $j$ is within the 2.8 x 2.8 km square centered at the centroid of wind project $r$, and let
\begin{equation}
    Q_{jr} = \int_0^{365}P_{jr}(t)dt
\end{equation}
be the occupancy time for eagles that winter in county $j$. Calculating this quantity gives managers concerned about maintaining populations that winter in county $j$ a clear estimate of the relative risk posed by different wind projects to birds that winter in their county.

\section{Results}
\begin{figure}[tbp]
    \centering
    \includegraphics[width = 5.61893in]{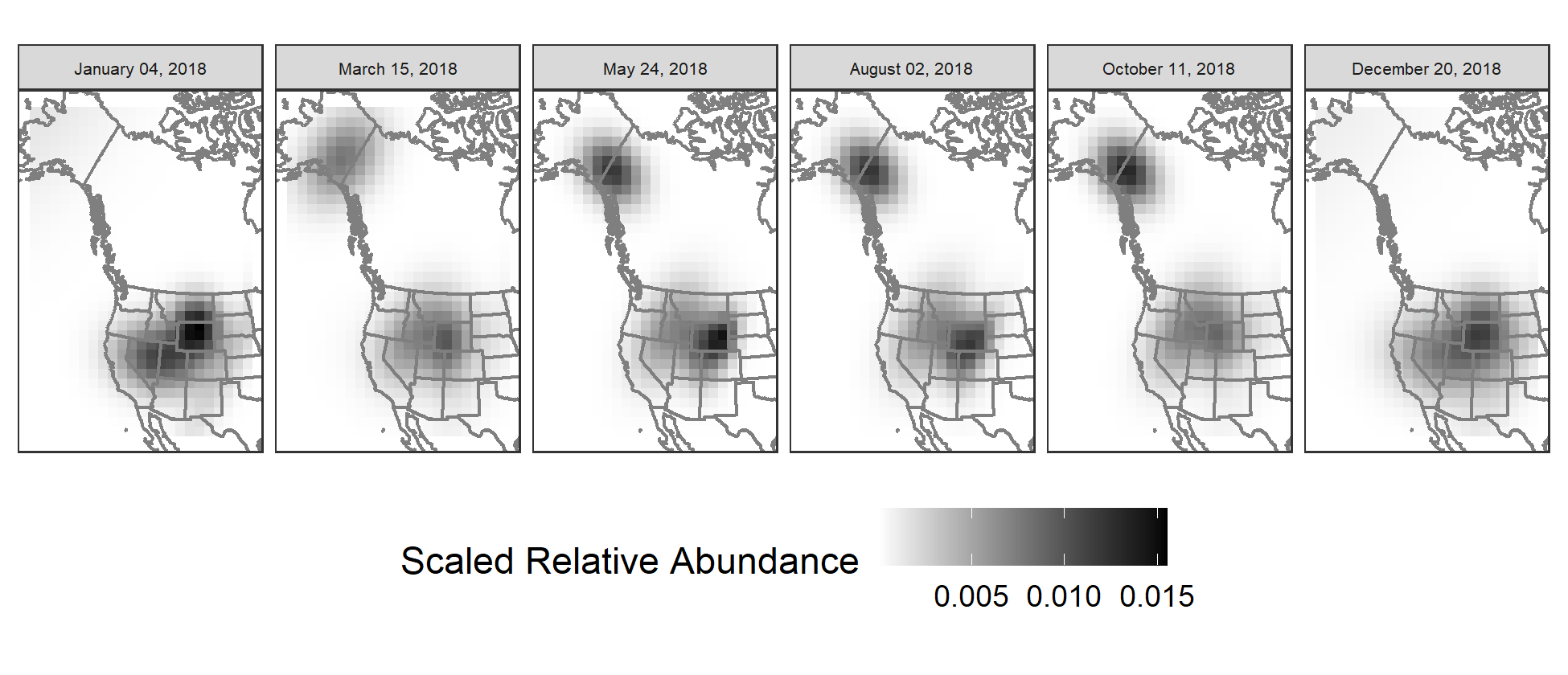}
    \caption{Posterior mean of the relative abundance of golden eagles in western North America for selected weeks.}
    \label{fig:estsddata}
\end{figure}

\begin{figure}[tbp]
    \centering
    \includegraphics[height = 3in]{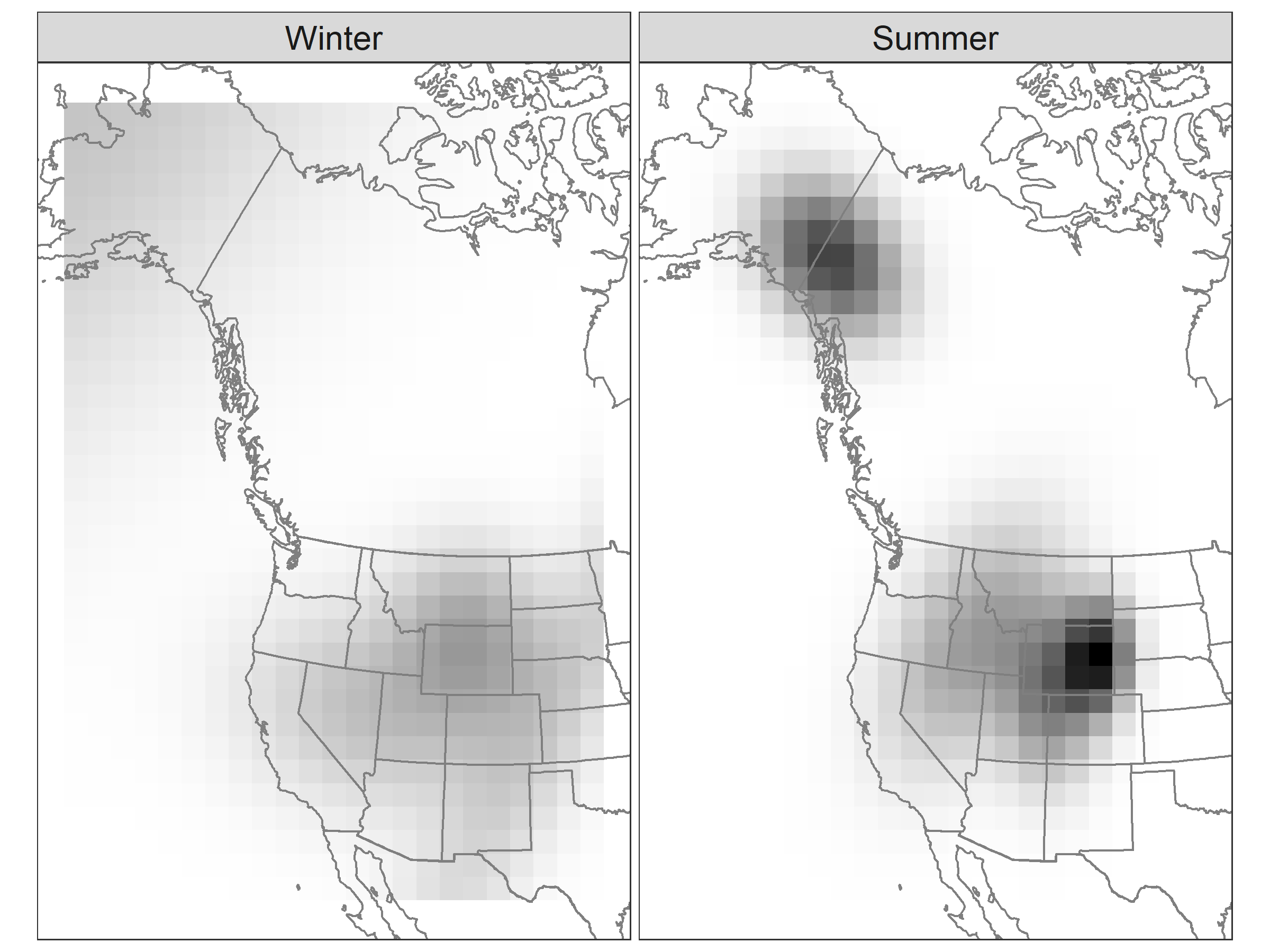}
    \caption{Posterior mean of the density of the stationary distribution of the OU process of the winter and summer attractive points.}
    \label{fig:allcenters}
\end{figure}

\begin{figure}[tbp]
    \centering
    \includegraphics[width = 5.61893in]{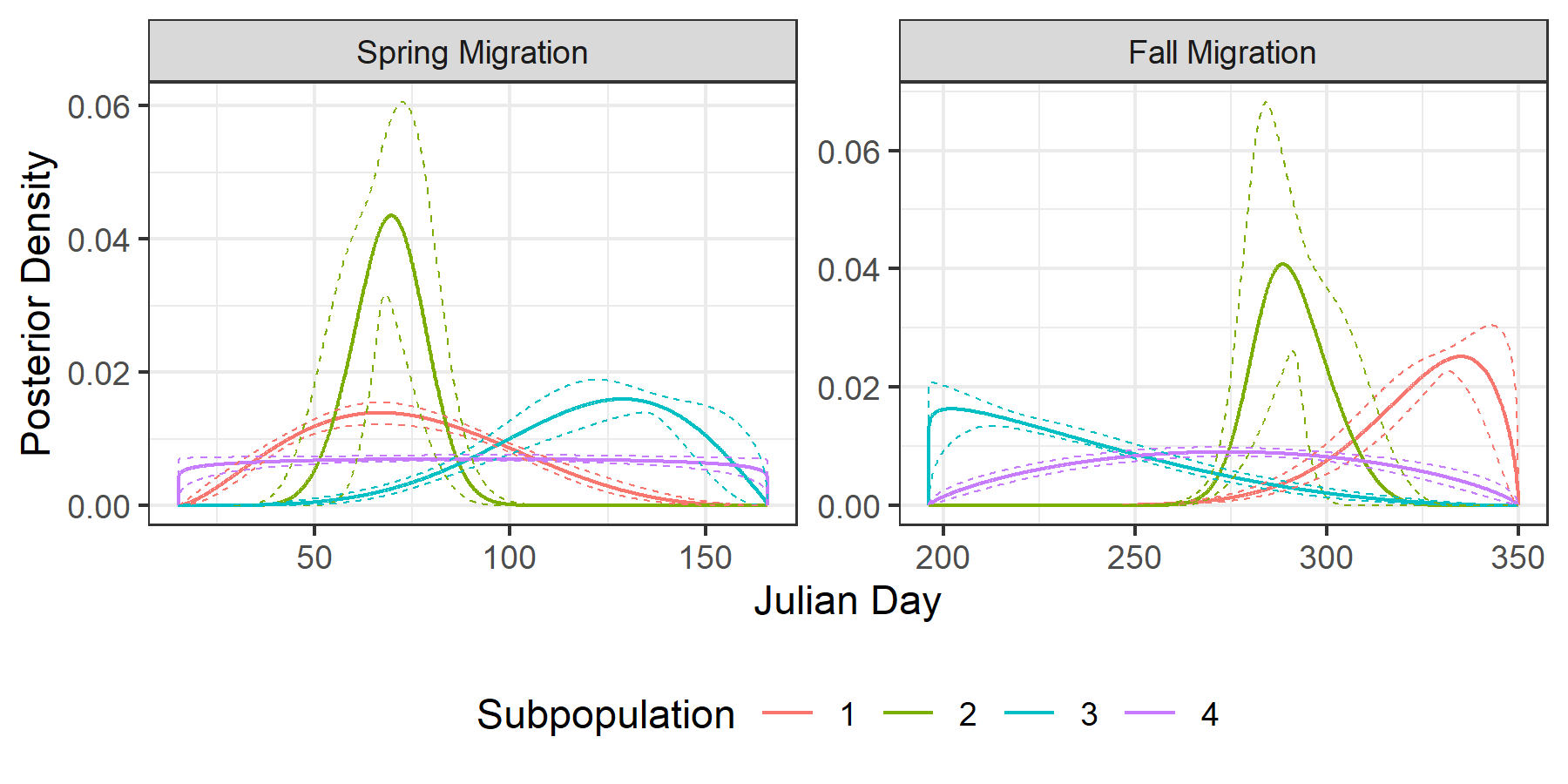}
    \caption{Posterior means and 95\% credible intervals for the distribution of the onset of spring and fall migration for each subpopulation.}
    \label{fig:springmig}
\end{figure}

We first show the results related to the movement dynamics of golden eagles in western North America. In Figure \ref{fig:estsddata} the posterior mean of the relative abundance, $\pi_t(\mathbf{x})$, closely matches the data in Figure \ref{fig:sdddata} \citep[for the posterior mean of the relative abundance for other weeks see Sections 2 and 7 of the Supplemental Material][]{FullYearOU_Supp}. We show the posterior mean of the distribution of first winter attractive points and the summer attractive points, $\mathbf{m}_{iy1}^{p}$ and $\mathbf{m}_{iy2}^{p}$, in Figure \ref{fig:allcenters}. The posterior mean for the distributions of the timings of the spring and fall migration, $b_{iy1}^{p}$ and $b_{iy2}^{p}$, with a 95\% credible interval are shown in Figure \ref{fig:springmig}. Posterior means and credible intervals for individual specific parameters (attractive points and starts of migration) can be found in Sections 1 and 6 of the Supplemental Material \citep{FullYearOU_Supp}.


\begin{figure}[tbp]
    \centering
    \includegraphics[width = 5.61893in]{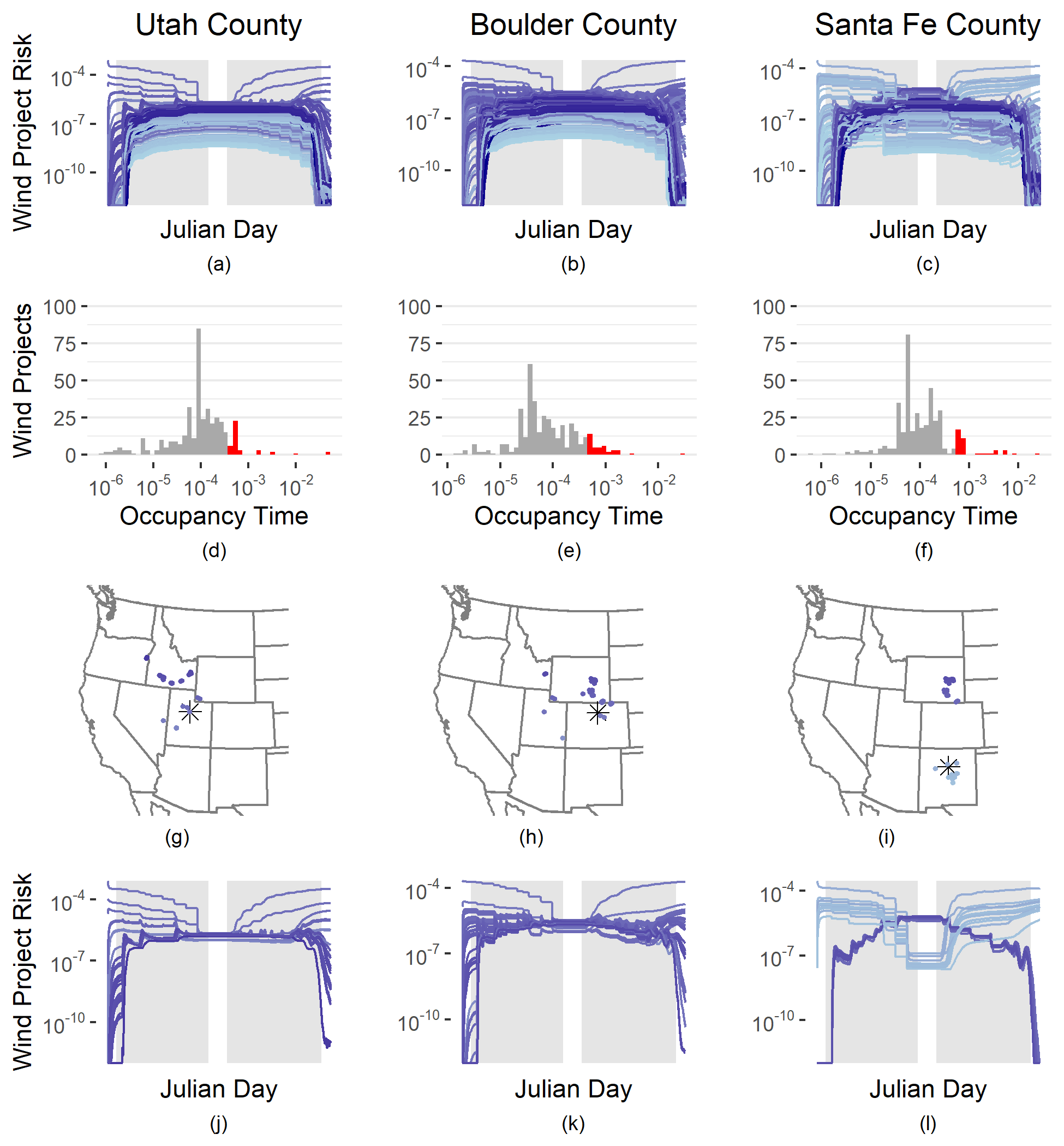}
    \caption{(a)-(c): Posterior mean wind project risk ($P_{jr}(t)$) for each wind project across the year for golden eagles that winter in Utah, Boulder, and Santa Fe counties; colors correspond to latitudinal location of the wind project, with darker blue for more northerly projects. (d)-(f): Histogram of the posterior mean of occupancy time ($Q_{jr}$) for each wind project; those wind projects with the greatest times are displayed in red (n = 40 projects). (g)-(i): The 40 wind projects with the highest posterior mean occupancy time and the centroid of the county (the black asterisk). (j)-(l): Posterior mean wind project risk for the 40 wind projects with the highest posterior mean occupancy time across the year. The axes for wind project risk and occupancy time are on a logarithmic scale.}
    \label{fig:risk}
\end{figure}

As described in Section \ref{sec:windmill}, we use our model to quantify the wind project risk for eagles that winter in select counties in the western United States \citep[for population level results see Section 4 in the Supplemental Material;][]{FullYearOU_Supp}. We estimate the posterior mean of $P_{jr}(t)$ (hereafter called wind project risk) for each of the wind projects. We show these posterior means for our three different counties of interest in Figures \ref{fig:risk}(a)-(c). Posterior means of the occupancy time, $Q_{jr}$, for each wind project and each county are shown in Figures \ref{fig:risk}(d)-(f) with the 40 wind projects with the highest posterior mean occupancy time highlighted in red. We then show the location of these 40 wind projects in Figures \ref{fig:risk}(g)-(i) and the posterior mean wind project risk for these selected projects in Figures \ref{fig:risk}(j)-(l).

We see that for wind projects close to the county of interest, risk to eagles starts high, falls, and then increases again (see the light blue in Figure \ref{fig:risk} (l)). For other wind projects we see relatively flat risk with spikes around the time when eagles wintering in each county migrate in the spring and fall (see the dark blue in Figure \ref{fig:risk} (j)). We also can see that wind projects that pose the greatest risk to eagles (as measured by posterior mean occupancy time) are generally close to the county of interest for Utah and Boulder counties; however, for individuals that winter in Santa Fe county our model indicates that the wind projects that have the highest risk are those wind projects which are close or in eastern Wyoming. These questions about subsets of the population could not be answered by using the eBird data alone but requires an estimate of the spatio-temporal dynamics of the entire population, as provided by the integrated model. This would enable managers to consider risk for subsets of the population that are relevant to their respective management area.

\begin{figure}[tbp]
    \centering
    \includegraphics[height = 3in]{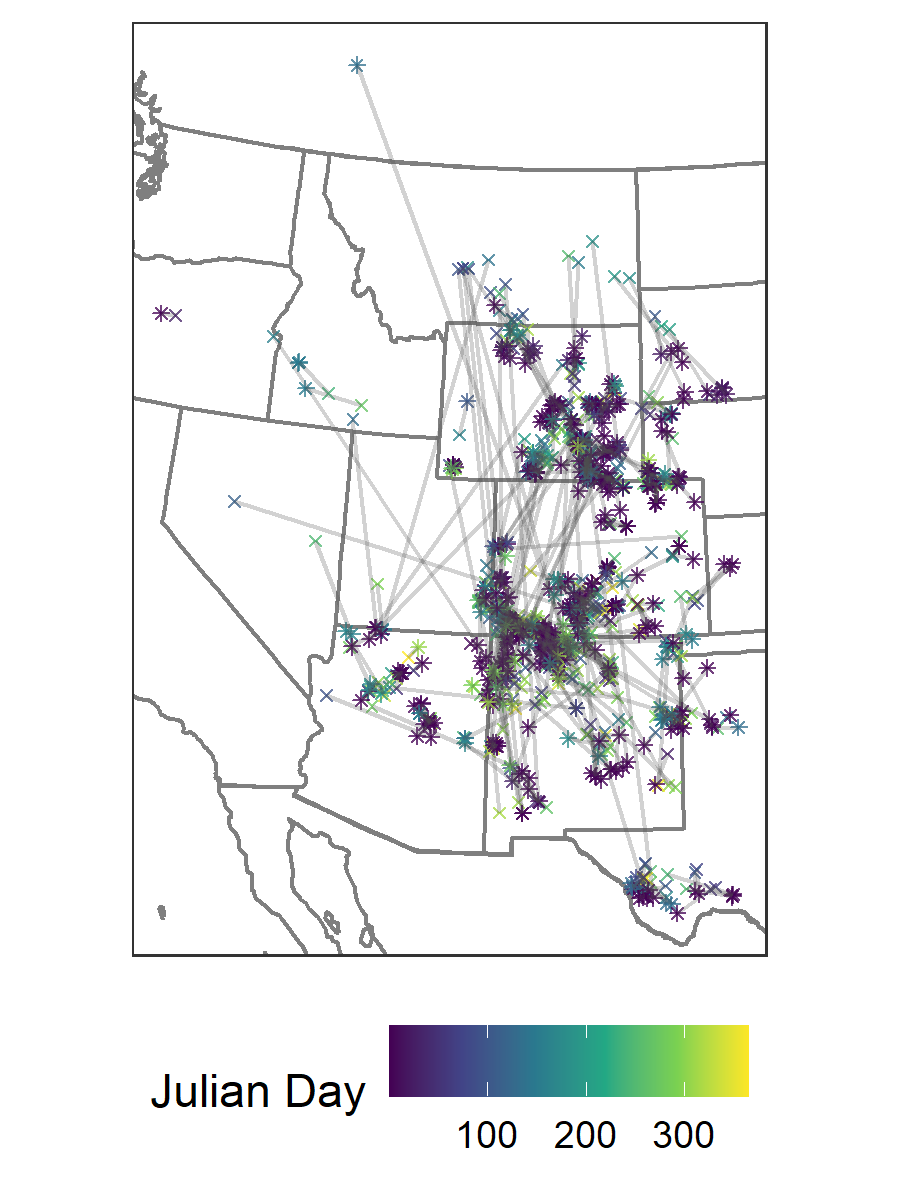}
    \caption{Selected validation telemetry data from data not included in our study. For each year we selected the first available telemetry reading (the asterisks) and the telemetry reading that was closest to one of the wind projects shown in Figure \ref{fig:windmills}(b) (the Xs). The color indicates the Julian day of the observation.}
    \label{fig:valdata}
\end{figure}

\begin{figure}[tbp]
    \centering
    \includegraphics[height = 3in]{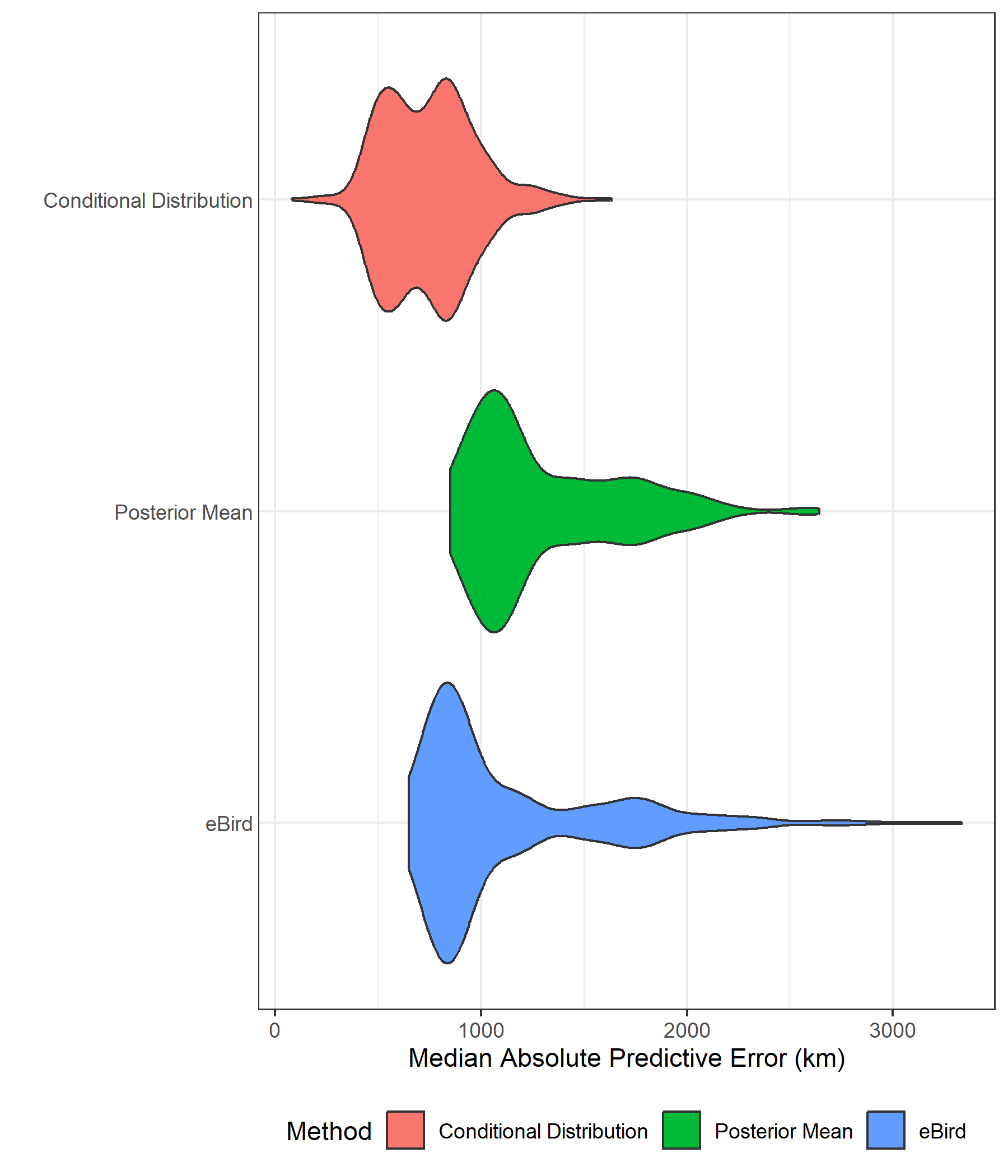}
    \caption{Violin plots for the median absolute prediction error for the three methods.}
    \label{fig:valcompare}
\end{figure}

We hold out the remaining 531 bird-years of incomplete telemetry data from 162 individual golden eagles (including the 93 individuals used in the study) for validation. We extract two observed telemetry data for each bird-year: the first observed telemetry data for each year, denoted $\mathbf{v}_{i1}$ which was observed at time $t_{i1}$, and the observed location that is closest to one of the wind projects from the data described in Section \ref{sec:data}, denoted $\mathbf{v}_{i2}$ which was observed at time $t_{i2}$. These observations are shown in Figure \ref{fig:valdata} \citep[the exact timing of the two telemetry observations for each year of telemetry data is given in Sections 5 and 8 of the Supplemental Material;][]{FullYearOU_Supp}. Our goal is to predict where a bird was at time $t_{i1}$ given that it is at location $\mathbf{v}_{i2}$ at time $t_{i2}$. Our motivation for this is inference on the possible winter range of a bird that was killed by a wind turbine at a known time and location. For each pair of validation locations we calculate a gridded estimate of the spatial distribution of the individual at time $t_{i1}$, conditional on being at location $\mathbf{v}_{i2}$ at time $t_{i2}$ by calculating
\begin{align}
    \varphi_i(\mathbf{y}) &= f(\mathbf{v}_{i1}|\mathbf{v}_{i2}) = \sum_{p=1}^P\frac{\Tilde{\pi}_{t_2}^p(\mathbf{v}_{i2})}{\sum_{l=1}^P\Tilde{\pi}_{t_2}^l(\mathbf{v}_{i2})}f^{p}(\mathbf{v}_{i1}|\mathbf{v}_{i2}) \\
    &\propto \sum_{p=1}^P \Tilde{\pi}_{t_2}^p(\mathbf{v}_{i2}) \frac{f^{p}(\mathbf{v}_{i2}|\mathbf{v}_{i1})\Tilde{\pi}_{t_1}^p(\mathbf{v}_{i1})}{\Tilde{\pi}_{t_2}^p(\mathbf{v}_{i2})} \nonumber \\
    &= \sum_{p=1}^P \Tilde{\pi}_{t_1}^p(\mathbf{v}_{i1}) \sum_{q=1}^Q f^{p}(\mathbf{v}_{i2}|\mathbf{v}_{i1}, b_{q1}^{p*}, b_{q2}^{p*}) \frac{g(b_{q1}^{p*}|\alpha_1^{p},\alpha_2^{p})g(b_{q2}^{p*}|\alpha_3^{p},\alpha_4^{p})}{\sum_{j=1}^Qg(b_{j1}^{p*}|\alpha_1^{p},\alpha_2^{p})g(b_{j2}^{p*}|\alpha_3^{p},\alpha_4^{p})} \nonumber
\end{align}
for every $\mathbf{y}\in\mathcal{D}$, where $f^{p}(\mathbf{v}_{i2}|\mathbf{v}_{i1}, b_{q1}^{p*}, b_{q2}^{p*})$ is the bivariate Gaussian density of \eqref{eq:OUintegrated},  $g(b_{q1}^{p*}|\alpha_1^{p},\alpha_2^{p})$ and $g(b_{q2}^{p*}|\alpha_3^{p},\alpha_4^{p})$ are the beta likelihoods of \eqref{eq:timingdist}, and $(b_{11}^{p*}, b_{12}^{p*}), \ldots, (b_{Q1}^{p*}, b_{Q2}^{p*})$ are as in Section \ref{sec:resultingPopDyn}. We then compute a median absolute predictive error (MAPE) for pair of validation locations by 
\begin{equation}
    \text{MAPE}_i(\text{Conditional Distribution}) = \sum_{\mathbf{y\in\mathcal{D}}} \hat{\varphi}_i(\mathbf{y}) \|\mathbf{y}-\mathbf{v}_{i1}\|_2
\end{equation}
where $\hat{\varphi}_i(\mathbf{y})$ is the posterior mean of $\varphi_i(\mathbf{y})$. For comparison we also consider computing the corresponding error if we use the posterior mean of the fitted relative abundance at time $t_{i1}$, $\pi_t(\mathbf{y})$,
\begin{equation}
    \text{MAPE}_i(\text{Posterior Mean}) = \sum_{\mathbf{y\in\mathcal{D}}} \hat{\pi}_{t_{i1}}(\mathbf{y}) \|\mathbf{y}-\mathbf{v}_{i1}\|_2,
\end{equation}
where $\hat{\pi}_{t_{i1}}(\mathbf{y})$ is the posterior mean of $\pi_{t_{i1}}(\mathbf{y})$, and also by using only the relative abundance eBird data for the week closest to $t_{i1}$, denoted by $\Tilde{t}_{i1}$
\begin{equation}
    \text{MAPE}_i(\text{eBird}) = \sum_{\mathbf{y\in\mathcal{D}}} s_{\Tilde{t}_{i1}}(\mathbf{y}) \|\mathbf{y}-\mathbf{v}_{i1}\|_2.
\end{equation}
Violin plots of the MAPE for each method are shown in Figure \ref{fig:valcompare} and numerical summaries are shown in Table \ref{tab:val_summary}. The median and mean are both lowest for the Conditional Distribution method, illustrating the increased predictive power we gain from pairing the eBird adaSTEM output with the telemetry data.

\begin{table}[tbp]
\centering
\begin{tabular}{l|rrr}
  \hline
Method & Median MAPE & Mean MAPE & Proportion Lowest MAPE \\ 
  \hline
Conditional Distribution & 748 & 747 & 84.4\% \\ 
  Posterior Mean & 1129 & 1304 & 0\% \\ 
  eBird & 897 & 1124 & 15.6\% \\ 
   \hline
\end{tabular}
\caption{Numerical summaries of validation error for the three methods.} 
\label{tab:val_summary}
\end{table}

\section{Discussion}
In this work we propose the first probabilistic full-year integrated movement model for jointly modeling individual tracking data and species distribution data. We use this model to assess wind project risk for golden eagles in the western United States at both population and subpopulation level. While assessing wind project risk at the population level can be done by using the eBird data alone, this joint model is needed for assessing risk for subsets of the population, such as those that winter in a specific location, which would be useful to obtain as wildlife management often occurs at the state level. We also use this model to predict a bird's location earlier in the year based on where it is at a later point; this showcases the usefulness of pairing the eBird species distribution data together with telemetry observations and could be used for predicting the spatio-temporal distribution of individuals given their location at a specified time. Such information could help support land use planning decisions that ultimately influence the conservation status of some migratory species of wildlife. Methods presented herein could be used in environmental impact assessments, specifically by estimating the spatio-temporal distribution of animals that could come into contact with a proposed project. For golden eagles specifically, results from the model could be included in Eagle Conservation Plans that are recommended by the U.S. Fish and Wildlife Service “…to help make wind energy facilities compatible with eagle conservation and the laws and regulations that protect eagles” \citep{USFWS_guidelines}.

In this work we took an empirical Bayes approach and fixed each eagle to one mixture component at the start. We could instead allow each eagle's individual telemetry data and parameters to follow a mixture model with no constraints on the weights. We could also increase the number of mixture components used to create a more flexible model, or use a Dirichlet process prior for a fully nonparametric model for heterogeneity in movement behavior; this could also include allowing mixture components that have no associated telemetry data, allowing insight into what portion of the eBird adaSTEM output is explained by population segments for which we do not have telemetry data. However, this would need to be done carefully to ensure that all mixture components still represent realistic eagle movement behavior.

The individual movement model utilized herein is flexible yet is simplistic in some ways. While this simplicity enabled efficient computation, it could be of interest to utilize a more realistic model for individual movement. We have a constraint in our model that $\sigma$ and $\beta$ are the same for all individuals in a subpopulation. This is rather restrictive and we could extend the model so that individuals have their own parameters that might even vary throughout the year, perhaps aligning with movement behavior (e.g. migration or central place foraging), to improve the fit of the model. However, such modifications would result in the loss of some of the analytic results that enable straightforward computation, and therefore would require efficient approximation of the dynamics for applications at the continental scale.

In movement ecology there is often a desire to link movement decisions to covariates, such as land cover \citep[e.g.][]{hanksContinuoustimeDiscretespaceModels2015, avgarIntegratedStepSelection2016} or weather \citep[e.g.][]{michelot2021varying}, and it would be of interest to include covariates in the IMM framework. Some possibilities for doing this would be to change our individual movement model into a commonly used movement model that includes covariates such as a step selection function or to modify our time-varying OU process to include covariates, possibly by allowing parameters in the model to be functions of land cover covariates and weather.

\section{Acknowledgments}
Robert K. Murphy is per contract with the U.S. Fish \& Wildlife Service—Division of Migratory Bird Management, National Raptor Program. We thank the eBird participants for their contributions, the eBird data management, and the U.S. Fish and Wildlife Service's Division of Migratory Bird Management, National Raptor Program for providing the Golden eagle location data. We also acknowledge Viviana Ruiz-Gutierrez, whose insights helped frame this paper.

This material uses data from the eBird Status and Trends Project at the Cornell Lab of Ornithology, eBird.org. Any opinions, findings, and conclusions or recommendations expressed in this material are those of the authors and do not necessarily reflect the views of the Cornell Lab of Ornithology.
\section{Funding}
Michael L. Shull and Ephraim M. Hanks were supported by the National Science Foundation (DMS-2015273, DBI-2433529). Frances E. Buderman was supported by the U.S. Department of Agriculture National Institute of Food and Agriculture and Hatch Appropriations under Hatch Project \#PEN04758 and Accession \#1024904 and the National Science Foundation (DBI-2433529).
\section{Disclosure Statement}
The authors report there are no competing interests to declare

\bibliography{bib.bib}
\bibliographystyle{apalike}

\end{document}